\documentclass[twocolumn,amsmath,amssymb]{jpsj3}
\usepackage{txfonts}
\usepackage{graphicx}% Include figure files
\usepackage{dcolumn}% Align table columns on decimal point
\usepackage{bm}% bold math
\usepackage{color}
\usepackage{accents}

\title{Robustness of Spin-Triplet Pairing and Singlet-Triplet Pairing Crossover in Superconductor/Ferromagnet Hybrids}

\author{Shiro Kawabata$^{1,2}$\thanks{E-mail: s-kawabata@aist.go.jp}, Yasuhiro Asano$^3$, Yukio Tanaka$^{4}$, and Alexander A. Golubov$^{5}$}
\inst{
$^1$Electronics and Photonics Research Institute (ESPRIT), National Institute of Advanced Industrial Science and Technology (AIST), Tsukuba, Ibaraki 305-8568, Japan
\\
$^2$Institut Laue-Langevin, 6 rue Jules Horowitz, BP 156, 38042 Grenoble, France
\\
$^3$Department of Applied Physics, Hokkaido University,
Sapporo 060-8628, Japan
\\
$^4$Department of Applied Physics, Nagoya University,
Nagoya 464-8603, Japan
\\
$^5$Faculty of Science and Technology and MESA+ Institute of Nanotechnology, University of Twente, 7500 AE, Enschede, The Netherlands
} %\\

\abst{
We have investigated the proximity effect in superconductor/ferromagnet junctions in a systematic manner to discuss the relationship between the zero-energy peak 
(ZEP) of the local density of states (LDOS) and  spin-triplet odd-frequency 
pairing.
By exactly solving the nonlinear Usadel equations, we have found that the ZEP is realized in a wide range of geometrical and material parameters in the case of the noncollinear magnetization.
This strongly suggests the robustness of the ZEP induced by spin-triplet odd-frequency pairing in such systems.
We also found that the crossover from singlet pairing to triplet pairing can be detected by measuring the F layer thickness dependence of the ZEP height.
Furthermore, we show how to observe signatures of spin-triplet odd-frequency pairing and the pairing crossover by LDOS measurement.
Our results provide a direct way to experimentally detect signatures of the odd-frequency pairing state.
}

\kword{odd frequency pairing, proximity effect, quasiclassical Green's function method, Usadel equation, spin triplet pairing, spintronics}

\begin{document}
\maketitle

\section{Introduction}

The study of the proximity effect in superconductor (S) - ferromagnet (F) hybrid structures has a long history
following the first theoretical proposal of the so-called $\pi$ state in a mesoscopic ring
containing SFS Josephson junctions.~\cite{Bulaevskii}.
The penetration depth of Cooper pairs into a diffusive normal metal (N) is characterized by
the length scale $\xi_T=\sqrt{\hbar D/2\pi T}$.
In a
ferromagnet, this length scale is considerably smaller and is given by $\xi_{h}=\sqrt{\hbar D/2E_\mathrm{ex}}$.
Here, $T$ is the temperature, $D$ is the diffusion constant, and $E_\mathrm{ex}$ is the
magnitude of the exchange potential in the ferromagnet.
Since the exchange field differently affects electrons with opposite spins,
 spin-singlet Cooper pairs are fragile under the exchange potential.
In addition to a small penetration length, the pairing function of spin-singlet pairs
spatially oscillates with changing sign
under the exchange potential,~\cite{ff,lo_jetp} which
enables the formation of $\pi$ states in SFS junctions.~\cite{buzdin_jetp}
Although the $\pi$ state was predicted theoretically in the 1970s, it has been confirmed experimentally
only recently.~\cite{ryazanov,kontos,robinson_prl}
Details of the progress of research on SF hybrids have been summarized in several review papers.~\cite{golubov_rmp,buzdin_rmp,bergeret_rmp}

Bergeret {\it et al.}~\cite{bergeret_prl} theoretically proposed a new type of
proximity-induced superconducting state in ferromagnets, the so-called long-range spin-triplet pairing state.
Inhomogeneous magnetic structures near the SF interface (see Fig.~1) rotate the spin direction of an electron, which induces
equal spin-triplet $s$-wave Cooper pairs in ferromagnets.~\cite{volkov,bergeret_rmp,eschrig_pt,kadigrobov_epl}
Such Cooper pairs should have an odd-frequency symmetry to satisfy the requirement of the Fermi-Dirac statistics of electrons.
Since equal spin-triplet pairs are not suppressed by the exchange potential, they have a
long-range length of penetration into ferromagnets
characterized by $\xi_T$.
Experimentally, such an effect has been observed first as the long-range
Josephson coupling in SFS junctions,~\cite{keizer_nature} where the ferromagnet is a half-metallic CrO$_2$ compound.
In clean half-metallic SFS junctions, equal-spin triplet pairs can have an odd-parity even-frequency symmetry.~\cite{eschrig_prl}
However, in the experimentally relevant dirty-limit case, the equal-spin triplet even-parity $s$-wave
odd-frequency symmetry is the only possible choice.
The experiment~\cite{keizer_nature} as well as anomalous conductance oscillations observed in SF hybrids~\cite{Sosnin} stimulated many
theoretical~\cite{asano_prl07,Asano2007,braude_prl,eschrig_natphys} and
experimental works~\cite{robinson_science,khaire,Sprungmann,anwar,Klose,Wang}. As a result,
a number of signatures of triplet correlations have been observed.
However, no experiment so far has shown direct and unequivocal evidence of long-range spin-triplet odd-frequency pairs.

%%%
%%%
%%%
%%%
\begin{figure}[b]
\begin{center}
\includegraphics[width=8.5cm]{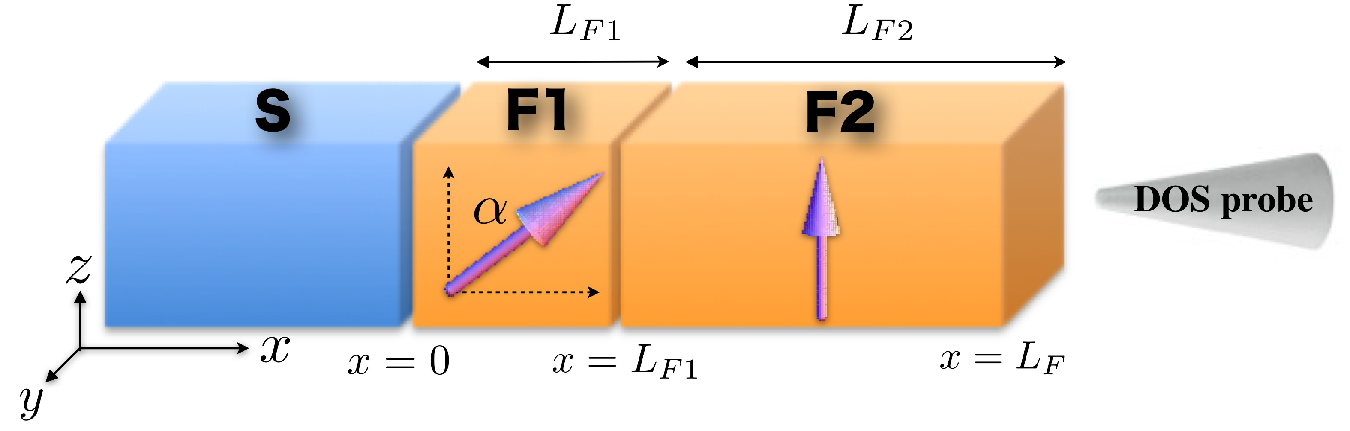}
\end{center}
\caption{(Color online) Model of an SF junction for observation of spin-triplet pairing through local density of states (LDOS) measurement.
The magnetization directions in F1 and F2 layers are collinear ($\alpha=0$) or noncollinear  ($\alpha=\pi/2$).
 }
\label{fig1}
\end{figure}
%%%
%%%
%%%
%%%

Existing theories predict that the presence of odd-frequency pairs causes
the enhancement of the zero-energy local density of states (LDOS).~\cite{bergeret_prl,asano_prl07,Asano2007,braude_prl,eschrig_natphys}
According to a number of theoretical studies of the proximity effect in various SF hybrid structures, {\it e.g.,} 
diffusive SF,~\cite{fominov1,houzet,cottet,linder3,Eschrig2009,Linder2010,Fominov2010,Trifunovic2010,Cottet11,Karminskaya,Ozaeta,Vasenko2013} 
clean SF,~\cite{halterman,Pugach,Trifunovic2011,Trifunovic,Wu,Melnikov}
S/magnetic-vortex,~\cite{Silaev,Zaikin} 
unconventional-superconductor/F,~\cite{yokoyama1,Sawa,Annunziata,Annunziata2012}  
and nonequilibrium SF junctions,~\cite{Bobkova} the relative fraction of odd-frequency pairs to even-frequency pairs depends sensitively on junction parameters such as the resistivity of F, the transparency of the SF interface, the amplitude of the exchange energy $E_\mathrm{ex}$, and the geometry of junctions.
To obtain clear evidence of spin-triplet odd-frequency pairs in experiments,
theoretical studies should show a way of optimizing the fraction
of odd-frequency spin-triplet pairs in a wide parameter range tunable in actual experiments.

In this study, by solving the full-spin Usadel equation~\cite{usadel,Belzig} in a wide parameter range, we systematically calculate 
the LDOS at a surface of a diffusive ferromagnet connected to a metallic superconductor.
In particular, we focus on the relationship between the magnitude of the zero-energy peak (ZEP) in the LDOS and the fraction of triplet odd-frequency pairs, and show the robustness of the presence of the ZEP.
Note that ZEP formation has been reported in measured tunneling conductance spectra in oxide-based SF heterostructures with nonuniform ferromagnets.~\cite{Dybko,Kalcheim1,Fridman,Visani,Kalcheim2}
However, the physical origin of ZEP in these structures is still unclear.
Therefore, we propose an experimental method of explicitly detecting the signature of spin-triplet pairing by measuring ZEP.

This paper is organized as follows.
In Sect.~2, we present a model of an SF junction and describe a numerical method of solving the nonlinear Usadel equation in such a system.
The numerical results of the LDOS for various parameters and the discussion of the robustness of ZEP are presented in Sect.~3.
In Sect.~4, the summary of our results is presented.
Throughout the paper, we confine ourselves to the regime of zero temperature and put $k_B=1$.

\section{Nonlinear Usadel Equation}
Let us consider SF junctions in the dirty limit  shown in Fig.~1. We assume that the junction is homogeneous within the $yz$ plane.
The magnetization in the ferromagnet can be either homogeneous or inhomogeneous. 
As an example of the inhomogeneous magnetization, we divide the F layer into two segments, F1 and F2, 
as shown in Fig.~1. In the F1 layer ($0 \leq x \leq L_{F1}$), the magnetic moment is in the $zx$ plane and rotated by $\alpha$ from the $z$-direction.
In the F2 layer ($L_{F1} \leq x \leq L_F$), on the other hand, the magnetic moment points the $z$-direction. The magnetic moments in the two layers 
are non-collinear to each other for $\alpha \neq 0$, which results in spin-flip scattering. 
The misorientation angle $\alpha$ in exchange-spring ferromagnets~\cite{Kneller,Fullerton,Gu} can be experimentally 
controlled by applying an in-plane magnetic field.
The length of the F2 layers is denoted by $L_{F2}$ (i.e., $L_F=L_{F1} + L_{F2}$).
The resistance due to the barrier at the SF interface is denoted by $R_B$ and that in the F layer by $R_N$.
In our calculation, we assume that the exchange energies $E_\mathrm{ex}$ in the F1 and F2 layers are the same, and that 
the interface between the two layers is transparent.

In the presence of spin-flip scattering, we have to solve a $4 \times 4$ matrix Usadel equation~\cite{usadel}
given by
\begin{align}
i\hbar D& \frac{d}{dx} \left( \check{g} \frac{d}{dx} \check{g} \right)- \left[ \check{H}, \check{g} \right]=0
,
\label{usadel0}
\end{align}
where $D$ is the diffusion constant in S and F.
The Hamiltonian $\check{H}$ and the Green function $\check{g}$ are respectively defined by
\begin{align}
\check{H}=& \left[ \begin{array}{cc}  \hat{E}(x,E)  & \hat{\Delta}(x,E)  \\
\undertilde{\hat{\Delta}} (x,E) & \undertilde{\hat{E}} (x,E) \end{array}\right]
,\\
\check{g}(x,E) =&
\left[ \begin{array}{cc} \hat{g}(x,E) & \hat{f}(x,E) \\
-\undertilde{\hat{f}}(x,E)& -\undertilde{\hat{g}}(x,E)
 \end{array}
 \right],
\end{align}
with
\begin{align}
\hat{E}(x,E) =& (E+i\delta) \hat{1} -\boldsymbol{V}(x)\cdot\hat{\boldsymbol{\sigma}}, \\
\undertilde{\hat{E}}(x,E)=& \left\{\hat{E}(x,-E)\right\}^\ast,\\
\undertilde{\hat{\Delta}}(x,E)=& \left\{\hat{\Delta}(x,-E)\right\}^\ast,
\end{align}
where $\delta$ is a small imaginary part indicating the retarded causality of the Green function.
Physically, $i\delta$ represents the self-energy due to inelastic scatterings, for instance, by the thermal excitation or electron-electron interactions.
Throughout this paper, we set $\delta/\Delta_0 =10^{-4}$, where $\Delta_0$ is the amplitude of the Pair potential at zero temperature.
Here, $\hat{\sigma}_j$ with $j=1, 2, 3$ are Pauli matrices and $\hat{\sigma}_0=\hat{1}$ is the $2\times 2$
unit matrix.
The magnetic moment $\boldsymbol{V}(x)$ in a ferromagnet is defined as
\begin{eqnarray}
\boldsymbol{V}(x)=
\left\{
\begin{array}{l}
E_\mathrm{ex}(\sin \alpha,0,\cos \alpha) \ \mathrm{for} \ \ \ \ 0 \le x \le L_{F1} \\
E_\mathrm{ex}(0,0,1) \  \ \ \ \ \ \ \ \ \ \mathrm{for} \ \ \ \ L_{F1} < x \le L_{F}
\end{array}
\right.
\label{mag}
.
\end{eqnarray}
Throughout this paper, $\check{\cdots}$ and $\hat{\cdots}$ indicate $4\times 4$ and $2\times 2$
matrices, respectively. In what follows, we only consider the $s$-wave spin-singlet pair potential
in a superconductor,
(i.e., $\hat{\Delta}=\Delta_0 i \hat{\sigma}_2$).
The particle-hole symmetry results in
\begin{align}
\undertilde{\hat{g}}(x,E) = \left\{\hat{g}(x,-E)\right\}^\ast,\\
\undertilde{\hat{f}}(x,E) = \left\{\hat{f}(x,-E)\right\}^\ast.
\end{align}

To solve Eq.~(\ref{usadel0}), we use the Riccati parameterization~\cite{Schopohl95,Schopohl98,Eschrig00,linder4} for the Green function, $i.e.,$
\begin{align}
\check{g}(x,E) =&
 \left[ \begin{array}{cc} \hat{N} & \hat{0} \\
{\hat{0}}& \undertilde{\hat{N}}
 \end{array}\right]
\left[ \begin{array}{cc} \hat{1} + \hat{\gamma} \undertilde{\hat{\gamma}} & 2\hat{\gamma} \\
-2\undertilde{\hat{\gamma}}& -( \hat{1} + \undertilde{\hat{\gamma}} \hat{\gamma})
 \end{array}\right],\label{ricatti}
 \end{align}
with
\begin{align}
 \hat{N} =& (\hat{1} - \hat{\gamma} \undertilde{\hat{\gamma}} )^{-1},\,\\
 \undertilde{\hat{N}} = & (\hat{1} - \undertilde{\hat{\gamma}}\hat{\gamma}  )^{-1}.
\end{align}
The normalization condition of the Green function is automatically
satisfied under the parameterization, $i.e.,$ $\check{g}\check{g}=\check{1}$.
The derivative of the inverse matrix $\partial_x\hat{N}$ can be obtained as 
\begin{align}
\partial_x\hat{N}=&\hat{N}\hat{A}\hat{N}
,
\label{derivative1}
 \end{align}
with
\begin{align}
\hat{A} =& (\partial_x\hat{\gamma}) \undertilde{\hat{\gamma}}
+ \hat{\gamma} \partial_x\undertilde{\hat{\gamma}},
\end{align}
which is obtained from the identity $\partial_x ( \hat{N} \hat{N}^{-1})=\hat{0}$.

Finally, the Usadel equation [Eq. (\ref{usadel0})] is reduced to two partial differential equations for $\hat{\gamma}$ and $\undertilde{\hat{\gamma}}$:
\begin{align}
&i\hbar D\left[ \partial_x^2 \hat{\gamma} + (\partial_x\hat{\gamma})\, \undertilde{\hat{f}}\, (\partial_x\hat{\gamma})
\right] - \hat{E} \hat{\gamma} + \hat{\gamma} \undertilde{\hat{E}}
+\hat{\Delta} - \hat{\gamma} \undertilde{\hat{\Delta}} \hat{\gamma}=0,
\label{usadel1}
\\
&-i\hbar D\left[ \partial_x^2 \undertilde{\hat{\gamma}} +(\partial_x\undertilde{\hat{\gamma}})
\, {\hat{f}}\,
(\partial_x\undertilde{\hat{\gamma}})
\right] - \undertilde{\hat{E}} \undertilde{\hat{\gamma}} + \undertilde{\hat{\gamma}} {\hat{E}}
+\undertilde{\hat{\Delta}} - \undertilde{\hat{\gamma}} {\hat{\Delta}} \undertilde{\hat{\gamma}}=0
\label{usadel2}
.
\end{align}
After taking the complex conjugation and $E\to -E$ in the above equations,
we find that $\hat{\gamma}^\ast(x,-E)$ and $\undertilde{\hat{\gamma}}(x,E)$
obey the same equation. Thus, we conclude that
\begin{align}
\undertilde{\hat{\gamma}}(x,E)= \hat{\gamma}^\ast(x,-E).
\end{align}

At the interface between a ferromagnet and a superconductor, the Kupriyanov-Lukichev
boundary condition~\cite{Kupriyanov} connects  Green functions on both sides, $i.e.,$
\begin{align}
2\Gamma_B \, \xi_{T_c} \, \check{g} \, \partial_x \check{g} =&\left[ \check{G}_S, \check{g}\right]
,
\label{kl0}
\end{align}
with $\Gamma_B=R_B/R_N^0$ and
\begin{align}
\check{G}_S=&\left[\begin{array}{cc}
{g}_s \hat{1} & f_s \hat{\sigma}_2 \\
f_s \hat{\sigma}_2 & -g_s \hat{1} \end{array}\right], \\
g_s =& \frac{E}{\sqrt{E^2-\Delta^2}},\\
f_s =& \frac{i\Delta}{\sqrt{E^2-\Delta^2}}
.
\end{align}
Here, $R_B$ is the resistance of the barrier at the SF interface and $R_N^0$ is the resistance of a ferromagnet whose length is $\xi_{T_c}=\sqrt{\hbar D/2\pi T_c}$, 
with $T_c$ being the superconducting transition temperature.
The resistance of the ferromagnet, $R_N$, is then given by $R_N^0 L_F / \xi_{T_c}$.
We obtain two boundary conditions:
\begin{align}
2\Gamma_B \,  \xi_{T_c} \, \partial_x \, \hat{\gamma} =& 2 g_s \, \hat{\gamma} + f_s ( \hat{\gamma} \, \hat{\sigma}_2 \, \hat{\gamma}
- \hat{\sigma}_2 ),
\label{kl1}
\\
2\Gamma_B \, \xi_{T_c} \, \partial_x \, \undertilde{\hat{\gamma}} =& 2 g_s\,  \undertilde{\hat{\gamma}}
- f_s ( \undertilde{\hat{\gamma}} \, \hat{\sigma}_2 \, \undertilde{\hat{\gamma}}
- \hat{\sigma}_2 ).
\label{kl2}
\end{align}
Since $g_s(-E)=g_s^\ast(E)$ and $f_s(-E)=f_s^\ast(E)$,
$\hat{\gamma}^\ast(x,-E)$ and $\undertilde{\hat{\gamma}}(x,E)$ satisfy
the same boundary condition.

By numerically solving the nonlinear differential equations [Eqs. (\ref{usadel1}) and (\ref{usadel2})] together with the boundary conditions [Eqs. (\ref{kl1}) and (\ref{kl2})], we 
calculate the LDOS as
\begin{align}
\frac{N(E)}{N_0}=\frac{1}{2}\mathrm{Tr} \left[ \mathrm{Re} \hat{g} (E) \right]
,
\end{align}
 and the pair function matrix as
\begin{align}
\hat{f} (E)
= &
\left[ f_0(E) \hat{1} + \vec{f} \cdot \hat{\boldsymbol{\sigma}} \right] \vec{\sigma}_2
\nonumber
\\
=&\left[\begin{array}{cc}
i f_1(E) +f_2(E)  &- i f_3(E)  - i f_0(E)  \\
- i f_3(E)  + i f_0(E)  &- i f_1(E)  +f_2(E)
 \end{array}\right],
 \end{align}
where $N_0$ is the normal DOS and $\vec{f} =(f_1,f_2,f_3)$.
The components $f_0$, $f_1$, $f_2$, and $f_3$ respectively represent the pairing function for the spin-singlet state
[$ \left( \left| \uparrow \downarrow \right\rangle -  \left| \downarrow \uparrow \right\rangle \right)/\sqrt{2}$],
 the equal-spin-triplet states [$ \left(\left| \uparrow \uparrow \right\rangle -  \left| \downarrow \downarrow \right\rangle\right) /\sqrt{2} $], 
 [$ \left( \left| \uparrow \uparrow \right\rangle +  \left| \downarrow \downarrow \right\rangle\right) /\sqrt{2} $)], and the opposite-spin-triplet state 
 [$ \left(  \left| \uparrow \downarrow \right\rangle + \left| \downarrow \uparrow \right\rangle \right)/\sqrt{2}$].
The LDOS $N(E)$ and the pairing function $f_i$ are local values depending on $x$.

In the dirty limit, the singlet component $f_0$ has an even-frequency symmetry, while the three triplet components $f_i$ ($i=1,2,3$) have an 
odd-frequency symmetry.
Note that the $y$ component of the magnetic moment $\boldsymbol{V}(x)$ is zero in this paper [see Eq.~(\ref{mag})].
Thus, the equal-spin-triplet component $f_2$ is always zero.

%%%%%%%%
%%%%%%%%
%%%%%%%%
%%%%%%%%
%%%%%%%%
\section{Results}
%%%%%%%%
%%%%%%%%
%%%%%%%%
%%%%%%%%
%%%%%%%%
In this section, we study the LDOS in the cases of uniform and nonuniform magnetizations systematically.
In order to show the robustness of the ZEP in the LDOS induced by spin-triplet odd-frequency pairs, 
we calculate the phase diagrams of the ZEP and the pair amplitudes as functions of several variables.
We also discuss how to detect signatures of long-range triplet pairing experimentally.

%%%
%%%
%%%
%%%
\begin{figure*}[t]
\begin{center}
\includegraphics[width=16.0cm]{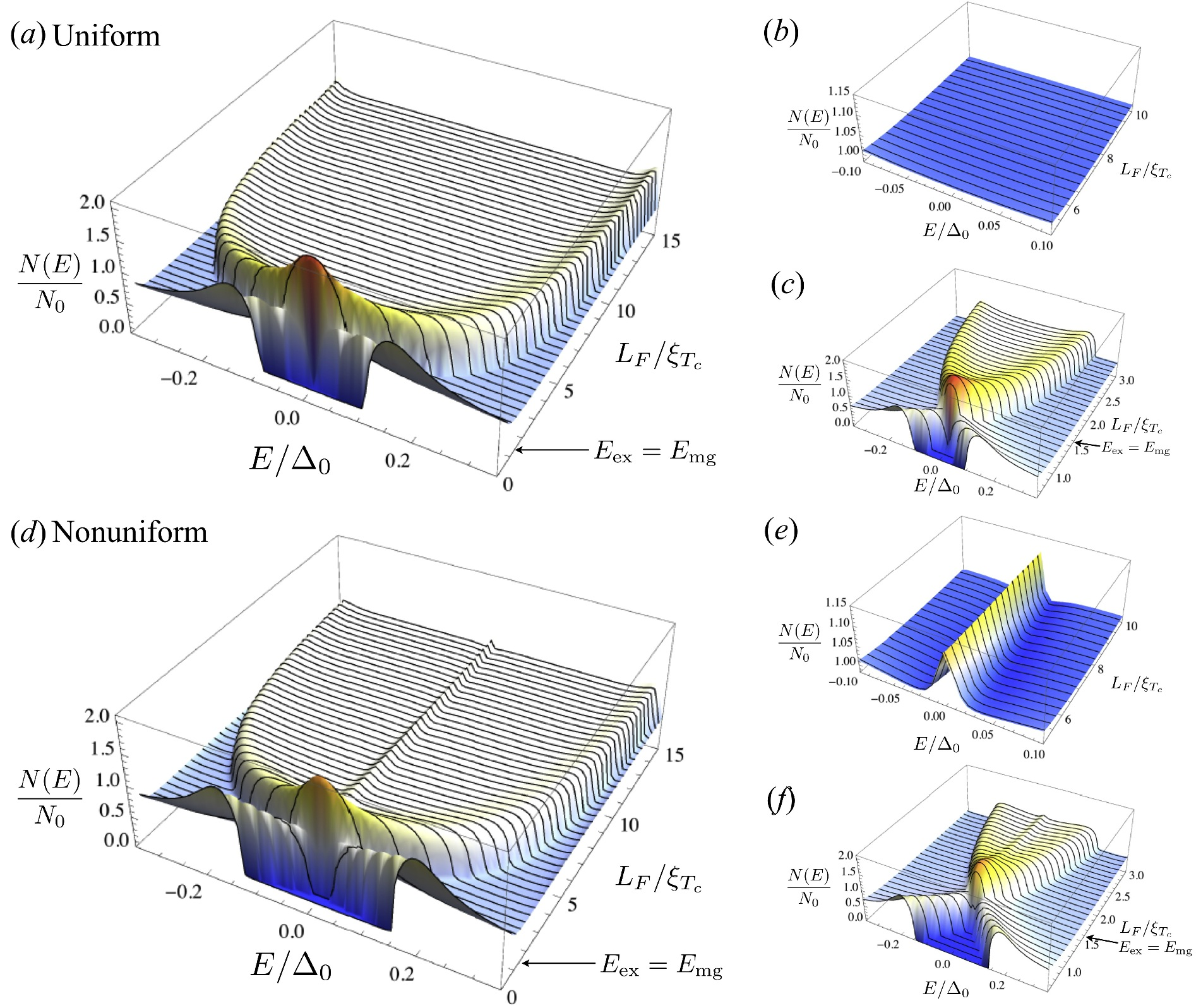}
\end{center}
\caption{(Color online) Local density of states (LDOS) $N(E)$ at the edge of the F$_2$ layer, $i. e.$, $x=L_F$, as a function of energy $E$ and the F layer thickness $L_F$ for an SF junction with (a) uniform ($\alpha=0$) and (d) nonuniform ($\alpha=\pi/2$) magnetizations with $L_{F1}=0.5  \xi_{T_c}$, $E_\mathrm{ex}/2 \pi T_c=0.1$, and $R_N^0/R_B=0.2$.
$N_0$, $\Delta_0$, and $\xi_{T_c}$ are the normal-state LDOS, superconducting gap, and coherence length at $T=T_c$, respectively.
Panels (b) and (e) are magnified images of (a) and (d) near $E=0$, respectively.
Panels (c) and (f) are magnified images of (a) and (d) around the small-$L_F$ regime, respectively.
The arrows indicate the resonant condition $E_\mathrm{ex}=E_\mathrm{mg}$, where $E_\mathrm{mg}$ is the minigap in the case of $E_\mathrm{ex}=0$.
 }
\label{fig2}
\end{figure*}
%%%
%%%
%%%
%%%

%%%%
%%%%
%%%%
%%%%
\subsection{Local density of states and pair functions}
%%%%
%%%%
%%%%
%%%%

Let us first discuss the LDOS in the uniform-magnetization case, $i. e.,$ $\alpha=0$.
 We consider a very weak ferromagnet by choosing $E_\mathrm{ex}/2 \pi T_c=0.1$ and a rather moderate proximity effect by 
setting the $R_N^0/R_B=0.2$.
In Figs.~2(a)-2(c), we show the dependence of the LDOS $N(E)$ at the edge of the F2 layer (i.e., at $x=L_F$) on
the F layer thickness $L_F$. 
When $L_F/\xi_{T_c}$ is much smaller than unity or when the Thouless energy $E_\mathrm{Th}$ is much larger than $E_\mathrm{ex}$, 
$i. e.,$ $E_\mathrm{Th}= \hbar D / L_F^2 \gg E_\mathrm{ex}$, a minigap is formed owing to the proximity effect.
In this case, we can neglect the effect of the magnetic moment on the proximity effect. 
As a result of the proximity effect, the minigap appears in LDOS, as shown in Figs.~2(a) and 2(c), for $L_F/\xi_{T_c} \ll 1$.
The magnitude of the minigap is approximately given by~\cite{golubov_rmp}
\begin{equation}
E_\mathrm{mg}
=
\frac{E_\mathrm{Th}}{1+\frac{R_B}{R_N}}
,
\end{equation}
as in the case of diffusive S/N(normal metal) junctions.
When we increase $L_F$ [$i.e.,$ decrease $E_\mathrm{Th} ( \sim L_F^{-2})$], the size of the minigap gradually decreases, as shown in Fig.~2(c).
Eventually, when the resonant condition $E_\mathrm{ex}=E_\mathrm{mg}$ is satisfied, the minigap is closed completely~\cite{Golubov2002}
and the ZEP is developed.
However, when we increase $L_F$ or decrease $E_\mathrm{Th}$ further, the LDOS profile near $E=0$ becomes almost flat because the minigap edges move outwards toward the superconducting-gap edge.
Therefore, the ZEP can be realized only near the resonant condition $E_\mathrm{ex}=E_\mathrm{mg}$.~\cite{Yokoyama2005,Yokoyama2006}
Note that, for a large $L_F \gg \xi_{T_c}$, the LDOS has peaks at $E= \pm E_\mathrm{ex} \approx \pm 0.35 \Delta_0$ for $E_\mathrm{ex} /2 \pi T_c =0.1$, which is in agreement with the theoretical prediction.~\cite{Vasenko2011}

To make the above points clearer, we plot the zero-energy density of state $N(0)$ and the amplitude of the pair functions $|f_i (0)|$ ($i=0,1,3$) as  
functions of $L_F$ in Fig.~3(a).
In the uniform-magnetization case, the long-range triplet components $f_1$ and $f_2$ are completely absent.~\cite{bergeret_rmp}
Near the resonant condition $E_\mathrm{ex}=E_\mathrm{mg}$, 
the short-range triplet component $f_3 (0)$ is rather more dominant than the singlet one $f_0 (0)$, as shown in the lower panel of Fig.~3(a). 
Two short-range components with a decay length $\xi_h \approx 2.2 \xi_{T_c}$ basically coexist with each other.
These observations are consistent with previous results.~\cite{Asano2007,yokoyama1}

In the case of a nonuniform magnetization ($\alpha=\pi/2$), on the other hand, 
the characteristic behaviors of $N(0)$ are largely different from those in the uniform-magnetization case owing to the appearance of the 
long-range spin-triplet components. The decay length of the long-range component is given by $\text{min}(\xi_T,\xi_\delta)$, where 
$\xi_{\delta}=\mathrm{Re} \left[ \sqrt{\hbar D/ i \delta } \right]$ is the decay length stemming from inelastic scattering. 
In the present results,
$\xi_\delta<\xi_T$ because we consider the limit of zero temperature.
The LDOS at the edge of the F2 layer is shown as a function of $L_F$ in Figs.~2(d)-2(f).
As clearly shown in Figs.~2(d) and 2(e), the ZEP develops not only near the resonant condition, but also in a wider range of $L_F$.

In order to understand the physical origin of the ZEP, we plot the zero-energy LDOS $N(0)$ and the pairing functions $f_i(0)$ at the edge of the F2 layer ($x=L_F$) 
as a function of $L_F$ in Fig.~3.
The results show that $N(0)$ is larger than $N_0$ (the ZEP develops) when
the amplitude of the long-range triplet component $f_1(0)$ is larger than those of the short-range components $f_0$ and $f_3$. 
The short-range components show exponentially damped oscillation as a function of $L_F$, whereas the long-range one decays considerably slowly with increasing $L_F$.
This is due to the fact that long-range pairs ($\left| \uparrow \uparrow \right\rangle, \left| \downarrow \downarrow \right\rangle$) have a 
zero center-of-mass momentum as in the case of singlet pairs in conventional SN junctions.
Therefore, in the case of the long F2 layer, {\it i.e.},  $L_{F2} \gg \xi_{h}$, one can have an almost pure long-range triplet component near the 
edge of the F${}_2$ layer.

Note that the dependences of $N(0)$ and $f_1(0)$ on $L_F$ in the regime $L_F/\xi_{h} \gg 1$ are closely related as shown in Fig.~3(b).
This can be explained as follows.
When only the long-range component $f_1(0)$ exists, the zero-energy LDOS is approximately given by
\begin{equation}
\frac{N(0)}{N_0}=\frac{1}{2} \mathrm{Tr}  \left[ \mathrm{Re} \hat{g}(0) \right] \approx  \sqrt{1 + \left| f_1(0) \right|^2}
,
\end{equation}
using the normalization condition
\begin{equation}
\hat{g}^2 + \hat{f} \undertilde{\hat{f}} =\hat{1}
.
\label{normalization}
\end{equation}
Thus, the $L_F$ dependence of ZEP is closely related to that of the long-range triplet odd-frequency component $f_1(0)$.~\cite{bergeret_rmp}
Therefore, we conclude that the systematic ZEP measurements by changing $L_F$ give strong evidence of the long-range spin-triplet correlations.
In Sect.~3.3, we will discuss in more detail ways to experimentally discriminate between the short- and long-range triplet components.

%%%%
%%%%
%%%%
%%%%
\subsection{Robustness of the zero-energy peak}
%%%%
%%%%
%%%%
%%%%

%%%
%%%
%%%
%%%
\begin{figure*}[t]
\begin{center}
\includegraphics[width=13.5cm]{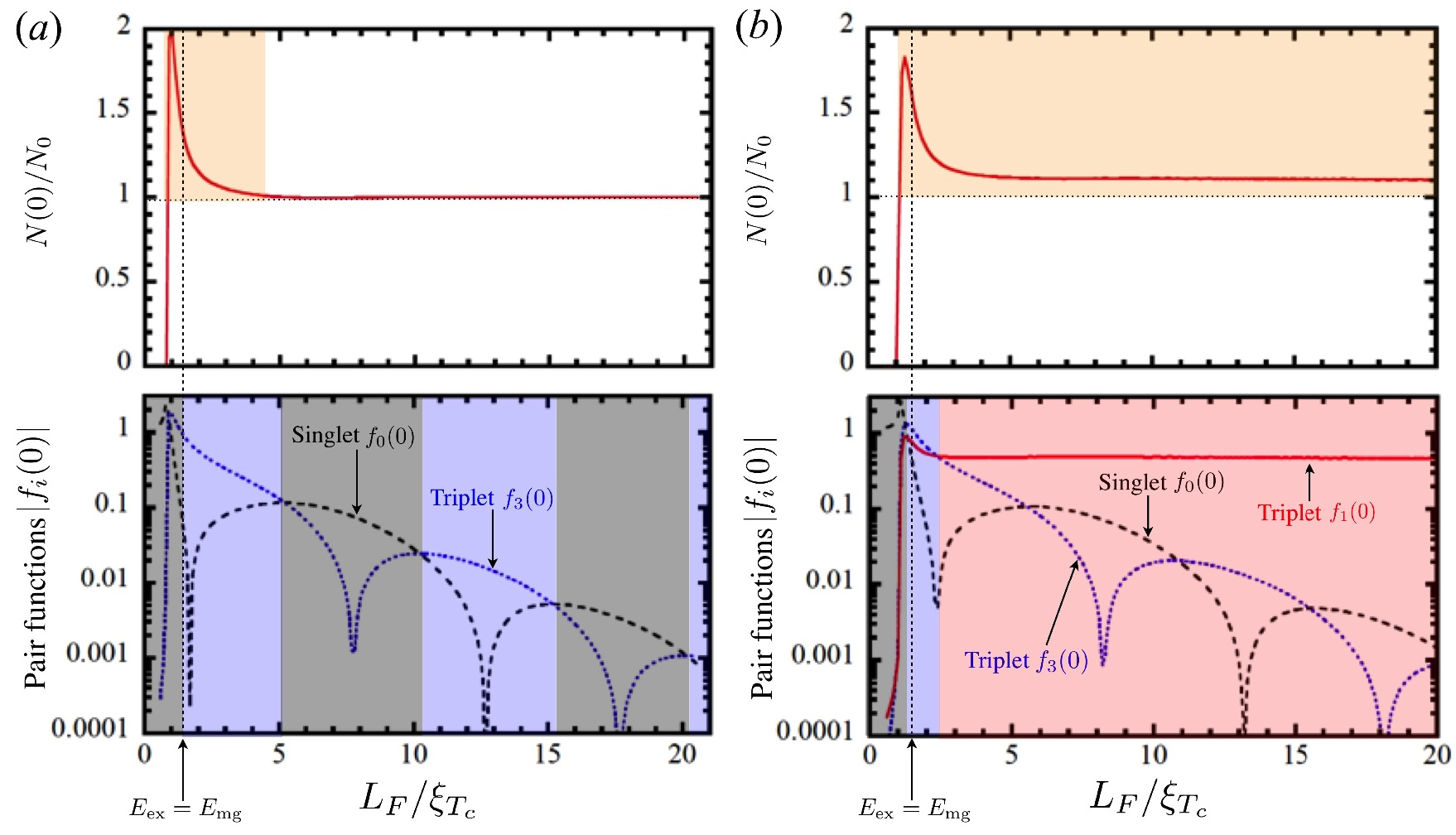}
\end{center}
\caption{(Color online) (a) Zero-energy LDOS $N(0)$ and the absolute value of the zero-energy pair-functions $f_i(0)$ as a function of the F layer thickness $L_F$ at the edge of the F2 layer ($x=L_F$). The results for the uniform magnetization $\alpha=0$ are in (a) and those for the nonuniform magnetization $\alpha=\pi/2$ are in (b).
$f_0$ (black dashed line), $f_3$ (blue dotted line), and $f_1$ (red solid line) are the short-range singlet, short-range triplet, and long-range triplet components, respectively.
The vertical dotted line corresponds to the resonant condition $E_\mathrm{ex}=E_\mathrm{g}$.
The parameters are $L_{F1}=0.5  \xi_{T_c}$, $E_\mathrm{ex}/2 \pi T_c=0.1$, and $R_N^0/R_B=0.2$.
 }
\label{fig3}
\end{figure*}
%%%
%%%
%%%
%%%
 %%%
%%%
%%%
%%%
%%%
\begin{figure*}[t]
\begin{center}
\includegraphics[width=10.0cm]{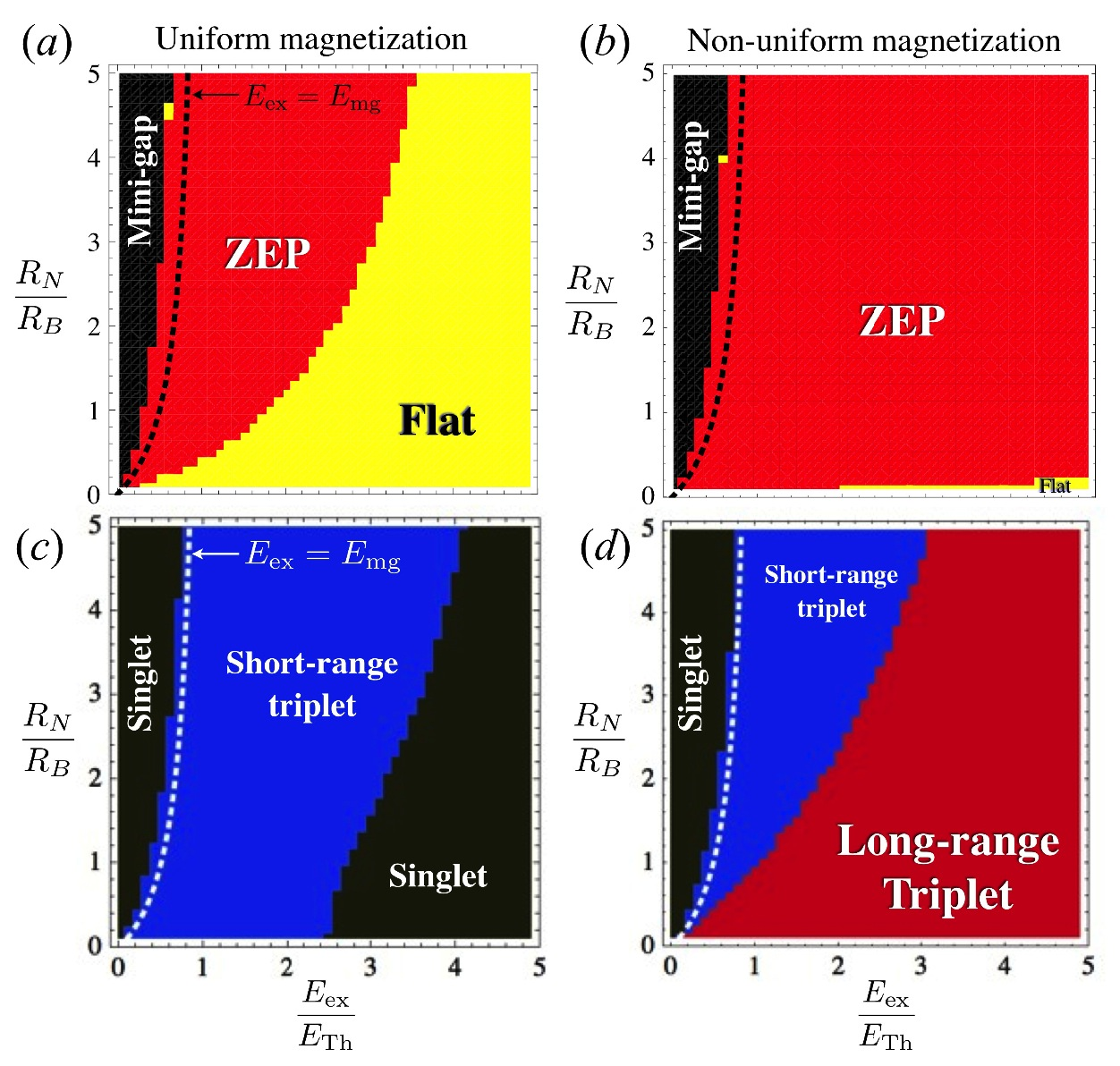}
\end{center}
\caption{(Color online) Phase diagram of the zero-energy LDOS $N(0)$ at $x=L_F$ as a function of $R_N/R_B$ and $E_\mathrm{ex}/E_\mathrm{Th}$ for an SF junction with (a)  uniform ($\alpha=0$) and (b) nonuniform ($\alpha=\pi/2$) magnetizations, where $L_{F1}=0.5  \xi_{T_c}$ and $L_{F}=4.0 \xi_{T_c}$.
The minigap, zero-energy peak (ZEP), and flat phase are respectively defined by the regions of $N(0)/N_0 < 0.98$,  $N(0)/N_0 >1.02$, and $0.98 \le N(0)/N_0 \le 1.02$.
Panels (c) and (d) show the phase diagram of the most dominant component of pair amplitudes at $x=L_F$ for an SF junction with uniform and nonuniform magnetizations, respectively.
Black, blue, and red regions are the singlet $f_0$-, short-range triplet $f_3$-, and long-range triplet $f_1$-dominant phases, respectively.
The dotted lines correspond to the resonant condition $E_\mathrm{ex}=E_\mathrm{mg}$.
 }
\label{fig4}
\end{figure*}
%%%
%%%
%%%
%%%
%%%

A number of theoretical papers have discussed the ZEP appearing in LDOS in a ferromagnet attached to a superconductor.
However, the investigation has been limited to very specific cases such as a very weak exchange field~\cite{Cottet11},
 a very strong exchange field (like half metals)~\cite{asano_prl07,braude_prl,eschrig_natphys}, a very small F layer thickness ($L_F \ll \xi_\mathrm{T_c}$)~\cite{braude_prl}, 
and a weak-proximity-effect regime (equivalently, $R_N^0/R_B \ll 1$).
Therefore, the natural question to ask is {\it how robust the presence of the ZEP induced by spin-triplet odd-frequency pairing is in actually?}
To answer this question, we calculate the zero-energy LDOS $N(0)$ and
the pair amplitudes $f_i(0)$ by systematically varying two parameters, (1) the exchange energy $E_\mathrm{ex}$ and (2) the barrier resistance $R_B$, which are controllable in experiments.
We will show the robustness of the presence of ZEP induced by spin-triplet odd-frequency pairs.

In Fig.~4, we show the phase diagram of $N(0)/N_0$ and the most dominant pairing function in F for uniform [Figs.~4(a) and 4(b)] and  
nonuniform [Figs.~4(c) and 4(d)] magnetization configurations as a function of $R_N/R_B$ and $E_\mathrm{ex}/E_\mathrm{Th}$.
In the calculation, we have assumed that $L_{F1}=0.5  \xi_{T_c}$ and $L_{F}=4.0 \xi_{T_c}$.
It is possible to define the following 
three phases: (i) minigap phase with $N(0)/N_0 \approx 0$, (ii) ZEP phase with $N(0)/N_0 >1$, and (iii) flat phase with $N(0)/N_0  \approx 1$.
In the calculation, we have defined the ZEP phase as regions with $N(0)/N_0 >1.02$ and the flat phase with as those with $0.98 \le N(0)/N_0 \le 1.02$ for practical convenience.
As was already discussed in Sect. 3.1, in the case of the uniform magnetization, only the short-range components $f_0$ and $f_3$ exist.
Thus, only in the vicinity of the resonant condition the ZEP phase develops [see the dotted line in Figs.~4(a) and 4(b)].

On the other hand, the ZEP phase in the nonuniform magnetization ($\alpha=\pi/2$) appears in wide parameter ranges 
of $E_\mathrm{ex}/E_\mathrm{Th}$ and $R_N/R_B$, as clearly shown in Fig.~4(b).
This can be attributed to the appearance of the long-range triplet component $f_1(0)$ [see Fig.~4(d)].
Note that the flat phase appearing for $R_N/R_B \ll 1$ and $E_\mathrm{ex}/E_{\mathrm{Th}}>2$ in Fig.~4(b) is due to the practical definition of the phase.
At $R_N/R_B \ll 1$, the proximity effect in the ferromagnet is very weak, which results in a very modest modulation of the LDOS. 
 Although the zero-energy LDOS in the flat phase is larger than $N_0$, it cannot be larger than $1.02 N_0$.
As a consequence, the flat phase appears in Fig. 4(b). 
From the above results, one can conclude that, in the SF junction with the nonuniform magnetization,  
{\it the appearance of the ZEP induced by odd-frequency spin-triplet pairs is very robust and insensitive to the device configuration and material parameters} 
as long as  $E_\mathrm{ex}/ E_\mathrm{mg} > 1$ is satisfied.
Therefore, the experimental observation of the ZEP gives unequivocal evidence of odd-frequency spin-triplet pairs.
 This is one of the important findings in this work.

%%%%
%%%%
%%%%
%%%%
\subsection{Zero-energy peak spectroscopy}
%%%%
%%%%
%%%%
%%%%

In this subsection, we study details of the ZEP structure in LDOS and propose an experimental method of detecting  odd-frequency spin-triplet pairs by analyzing 
the ZEP, $i.e.,$ {\it the ZEP spectroscopy}.
The possibility of observing {\it the singlet-to-triplet crossover} by ZEP spectroscopy is discussed as well.

We study the deviation of the LDOS at the zero-energy $N(0)$ from its normal value $N_0$, {\it i.~e.,} $ \left|  \delta \nu_0  \right| \equiv \left| N(0)/N_0 -1\right|$.
The LDOS is calculated at the end of the F2 layer ($x=L_F$).
In the calculation, we fix $L_{F1}=0.5  \xi_{T_c}$ and $E_\mathrm{ex}/2 \pi T_c=0.1$ as in Fig.~3.
In Fig.~5, we show  $|\delta \nu_0|$ as a function of $L_F$ in the cases of a moderate-proximity-effect regime ($R_N^0/R_B=0.2$) and a strong-proximity-effect regime ($R_N^0/R_B=1.0$).
In the strong-proximity-effect regime, the widely used linearized Usadel approach is not justified at all.
In the case of the uniform magnetization ($\alpha=0$), $|\delta \nu_0|$ shows the oscillatory damped behavior, as shown by the thick broken line in Fig.~5.
This means that $\delta \nu_0$ changes its sign almost periodically with the an increase in $L_F$.
Such behavior is consistent with previous theoretical predictions~\cite{Vasenko2011,Buzdin00,Vasenko} as well as with experimental results.~\cite{kontos2,Boden}
%%%
%%%
%%%
%%%
\begin{figure}[t]
\begin{center}
\includegraphics[width=8.5cm]{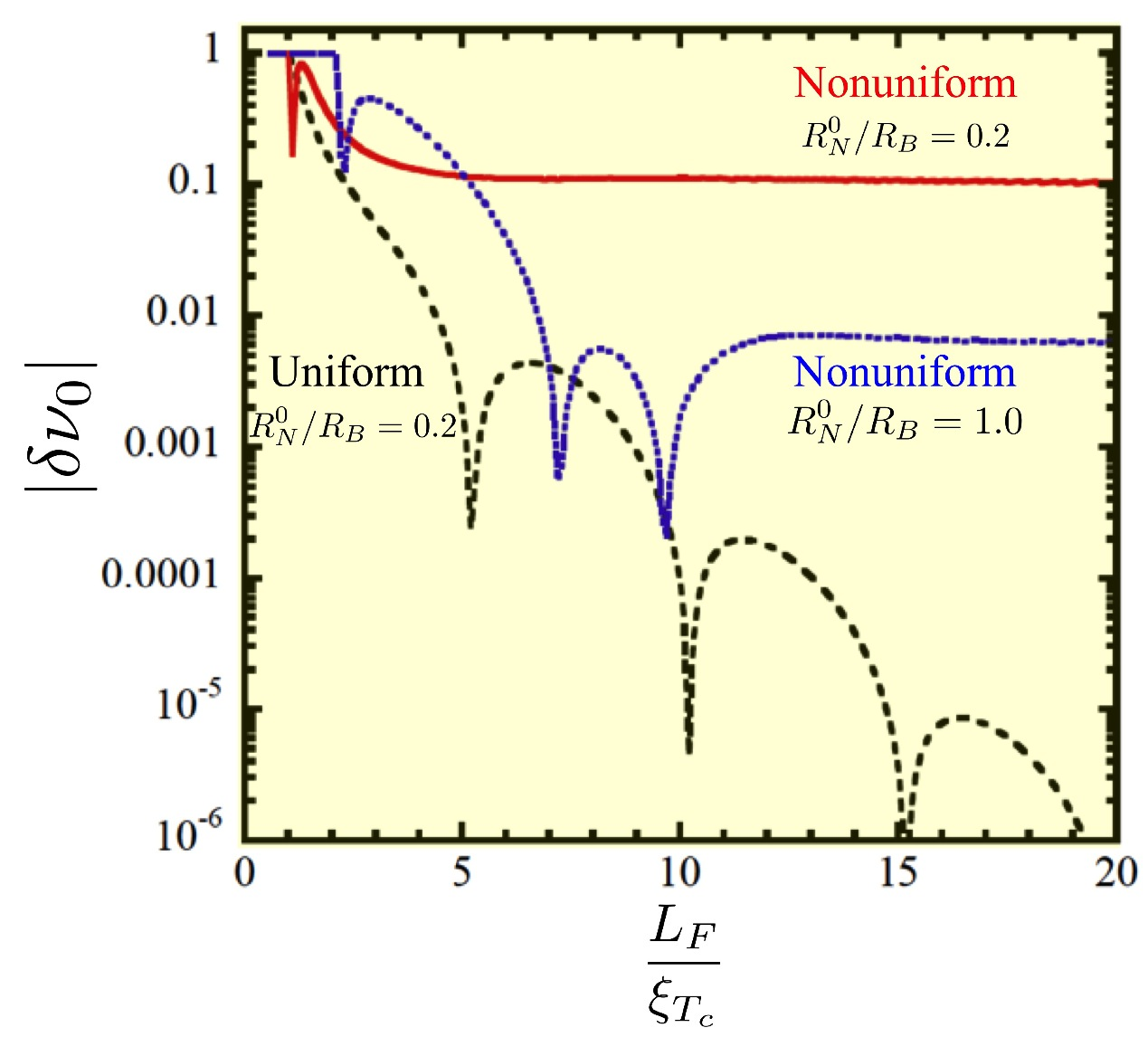}
\end{center}
\caption{(Color online) Zero-energy peak spectroscopy.
Deviation of the zero energy LDOS $N(0)$ from its normal value $N_0$, {\it i.e.,} $\delta \nu_0 = N(0)/N_0 -1$, as a function of the F layer thickness $L_F$ for an SF junction with $\alpha=0$ (black dashed line) and $\alpha=\pi/2$ (red solid and blue dotted lines) for different values of $R_N^0/R_B$.
The LDOS is evaluated at $x=L_F$.
The parameters are $L_{F1}=0.5  \xi_{T_c}$ and $E_\mathrm{ex}/2 \pi T_c=0.1$.
 }
\label{fig5}
\end{figure}
%%%
%%%
%%%
%%%
%%%
%%%
%%%
%%%
\begin{figure}[t]
\begin{center}
\includegraphics[width=8.5cm]{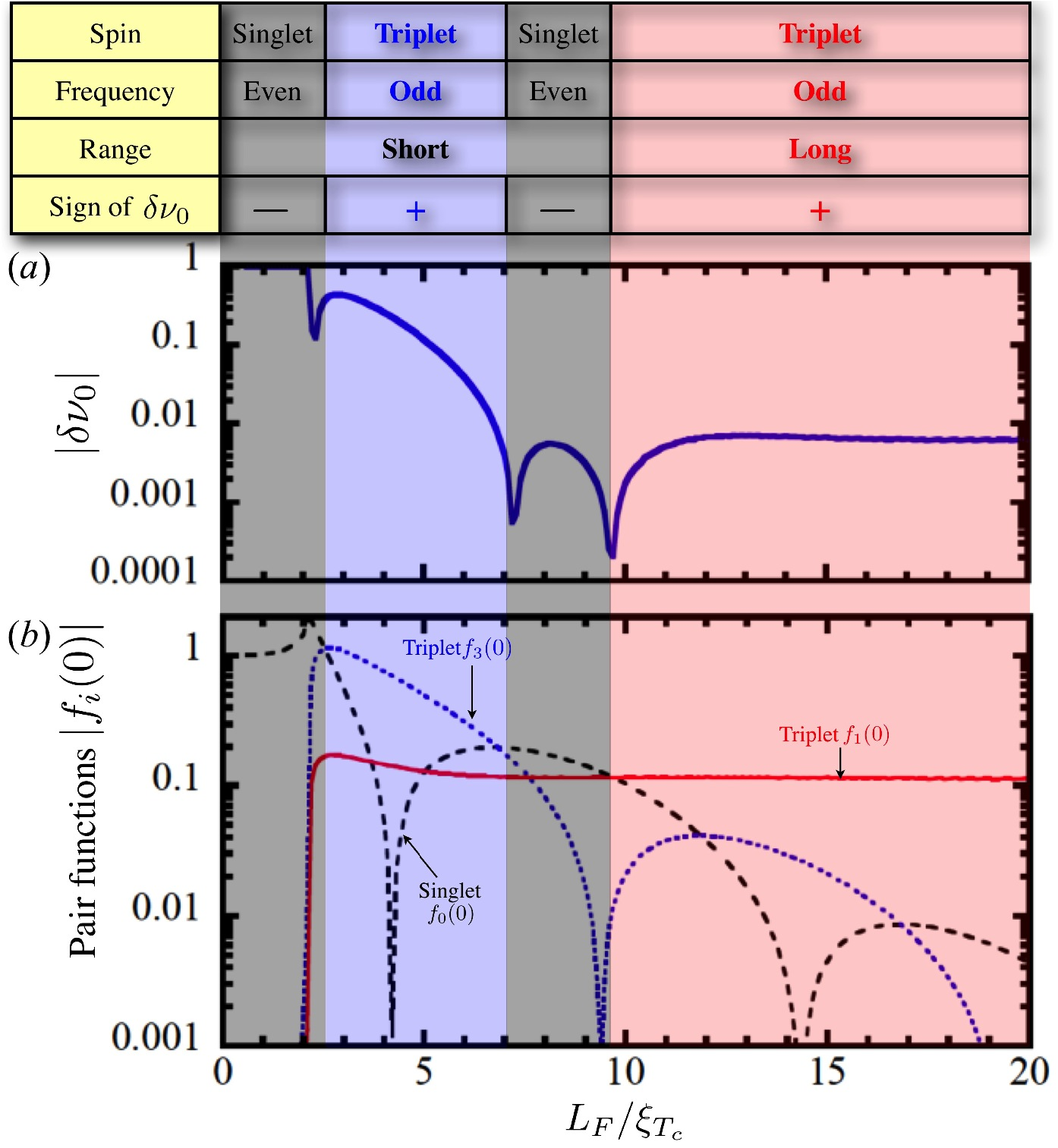}
\end{center}
\caption{(Color online) (a) Deviation of the zero-energy LDOS $N(0)$ from its normal value $N_0$, {\it i.e.,} $ \left| \delta \nu_0 \right| = \left| N(0)/N_0 -1 \right|$ at $x=L_F$ as a function of the F layer thickness $L_F$ for an SF junction with a nonuniform magnetization ($\alpha=\pi/2$) in a strong proximity regime ($R_N^0/R_B=1.0$).
Panel (b) shows the absolute value of  pair functions $f_i(0)$ at $x=L_F$ as a function of $L_F$.
$f_0$ (black dashed line), $f_3$ (blue dotted line), and $f_1$ (red solid line) are the short-range singlet, short-range triplet, and long-range triplet components, respectively.
The parameters are $L_{F1}=0.5  \xi_{T_c}$ and $E_\mathrm{ex}/2 \pi T_c=0.1$.
The gray, blue, and red region  correspond to the singlet-, short-range triplet-, and long-range triplet-dominant phases.
The upper table shows pairing symmetries for the most dominant component and the signs of $ \delta \nu_0$.
 }
\label{fig6}
\end{figure}
%%%
%%%
%%%
%%%

In contrast, the behaviors of $|\delta \nu_0|$ for the inhomogeneous magnetization ($\alpha=\pi/2$) are largely different from those for the uniform one.
Owing to the development of the long-range triplet component $f_1(0)$, $|\delta \nu_0|$  decays very slowly as a function of $L_F$, as shown by the solid and dotted 
lines in Fig.~5.
The sign of $\delta \nu_0$ is always $positive$ as long as $E_\mathrm{ex} > E_\mathrm{mg}$ [see also Fig.~3(b)].
The above results suggest that it is possible to distinguish between the spin-singlet even-frequency component ($f_0$) and spin-triplet odd-frequency 
ones ($f_1$, $f_2$ and $f_3$) by systematically measuring the LDOS at zero energy as a function of $L_F$.
Namely, when the even- and odd-frequency pair dominant phases crossover with each other, the crossover points correspond to the deep minima of $|\delta\nu_0|$.

To confirm our prediction, we also calculate the pairing functions as a function of $L_F$ in Fig.~6, 
where the amplitudes of the pairing functions are shown in (b) with the results of $|\delta \nu_0|$ with $R^0_N/R_B=1$ in (a). 
The dip positions of $|\delta \nu_0|$ are almost identical to the crossover points between the even- and odd-frequency pair dominant phases.
The physical origin of the above remarkable phenomenon can be explained as follows.
When all the components $f_i$($i=0-3$) coexist, $\delta \nu_0$ can be expressed by Eq. (\ref{normalization}) 
as~\cite{bergeret_rmp}
\begin{equation}
\delta \nu_0=
\frac{N(0)}{N_0} -1
\approx
-\frac{\left|f_0(0) \right|^2 }{2} + \sum_{i=1,2,3} \frac{ \left| f_i(0) \right|^2}{2} 
,
\end{equation}
by assuming $|f_i| \ll 1$, and considering the facts that $\mathrm{Im} f_0 (0)=0$ for the singlet component and $\mathrm{Re} f_i (0)=0$ for the triplet components ($i=1,2,3$).~\cite{bergeret_rmp,Silaev}
Therefore, the spin-singlet even- and spin-triplet odd-frequency components respectively have negative
and positive contributions to $\delta \nu_0$.
Thus, $\delta \nu_0$ changes its sign at the crossover points in $L_F$.
This gives rise to a dip structure in the $ |\delta \nu_0 |$ vs $L_F$ curve shown in Fig.~6(a).

The even-odd frequency (singlet-triplet) crossover happens even in
the weak- or moderate-proximity-effect regime ($R_N^0/R_B <1$).
As shown in Fig.~3(b) in the moderate-proximity-effect regime, the amplitude of the long-range component $f_1(0)$ is rather larger than that in the 
strong proximity cases in Fig.~6(b).
Therefore, the crossover (from short-range singlet to long-range triplet) happens only once near the resonant condition.
By extending the argument to the $uniform$ SF junction shown in Figs.~3(a) and 5,
the well-known zero-energy LDOS oscillations can be
interpreted as a result of a series of (short-range) singlet-to-(short-range) triplet crossovers.
Therefore, {\it  it is possible to identify the crossover points from the dip positions of $|\delta \nu_0|$}.
This remarkable feature has never been observed in previous studies and is one important finding of this study.
From the above results, it is clear that the systematic LDOS measurement by changing the exchange 
field $E_\mathrm{ex}$, the F layer thickness $L_F$, and the barrier resistance $R_B$ gives  
unequivocal evidence of the novel long-range triplet odd-frequency pairs.

%%%
%%%
%%%
%%%
\begin{figure}[t]
\begin{center}
\includegraphics[width=8.5cm]{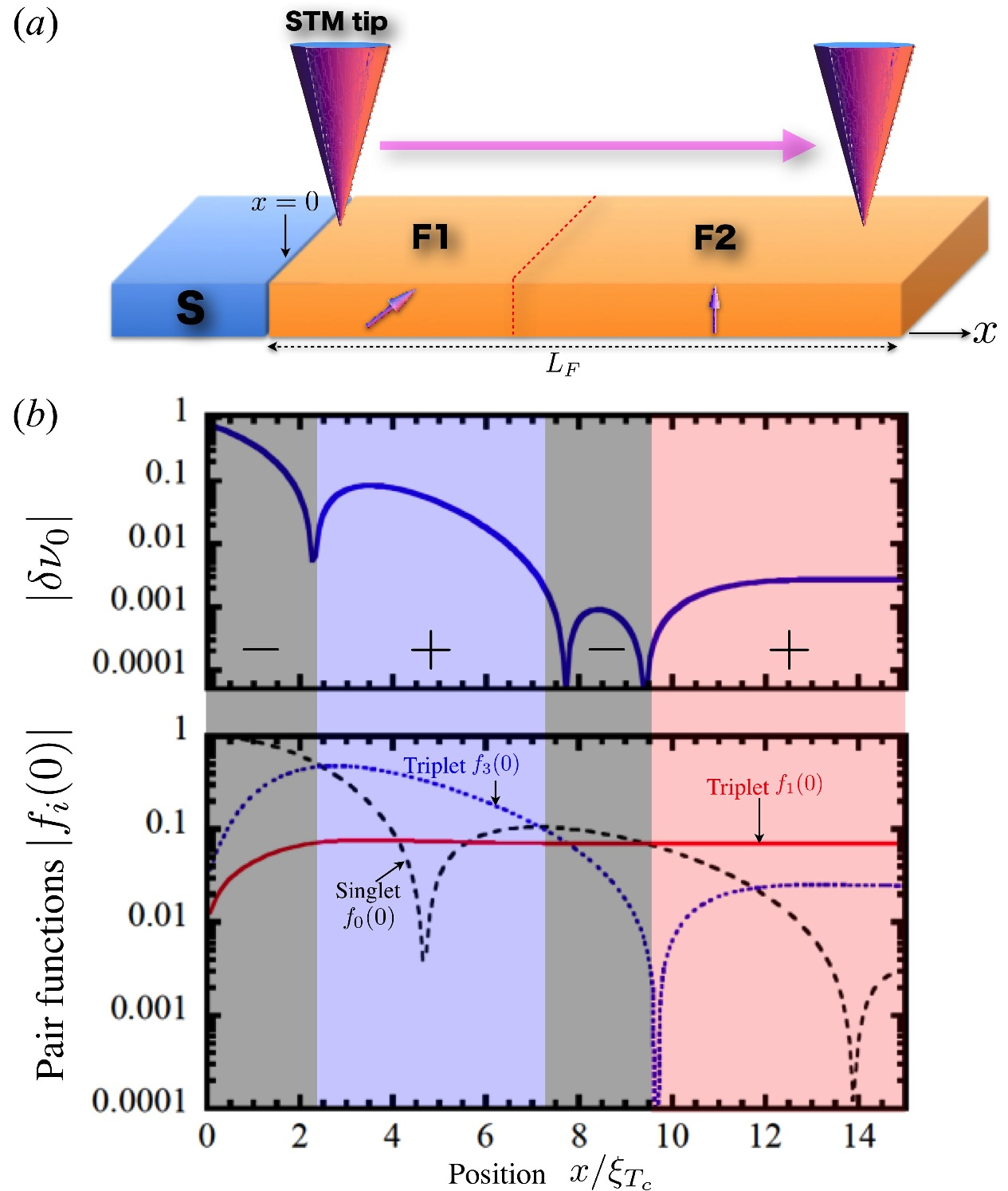}
\end{center}
\caption{(Color online) (a) Scheme of an inhomogeneous SF junction in contact with an STM tip.
Panel (b) shows the position $x$ dependence of $ \delta \nu_0  (0)$ and the pair functions $f_i(0)$ ($i=0,1,3$) in the F layer, where $\pm$ corresponds to the sign of $\delta \nu_0$.
$f_0$ (black dashed line), $f_3$ (blue dotted line), and $f_1$ (red solid line) are the short-range singlet, short-range triplet, and long-range triplet components, respectively.
The parameters are $\alpha=\pi/2$, $L_{F1}=0.5  \xi_{T_c}$, $L_{F2}=14.5  \xi_{T_c}$, $R_N^0/R_B=2.0$, and $E_\mathrm{ex}/2 \pi T_c=0.1$.
}
\label{fig7}
\end{figure}
%%%
%%%
%%%
%%%

Finally, we would like to propose an experimental method of detecting the signature of the odd-frequency pairing and  observing the singlet-to-triplet crossover.
Figure 7(a) shows a scheme of an SF junction in contact with a scanning tunneling microscope (STM) tip for measuring the differential conductance or the LDOS in the F layer.
The spatial dependence of the LDOS of F in the strong proximity regime is plotted in Fig.~7(b).
As clearly shown in Fig.~7(b), the measurement of the position dependence of the zero-energy LDOS enables clear identification of the 
long-range odd-frequency pairing as well as of the the singlet-to-triplet crossover.

The characteristic behaviors of $|\delta \nu_0|$ and $|f_i(0)|$ for $x \gg \xi_{T_c}$ in Fig.~7(b) are very similar to those 
for $L_F \gg \xi_{T_c}$ in Fig.~6. The physics happening at the edge of a sufficiently long ferromagnet and that at a point far enough from the 
SF interface should be the same.
However, the behaviors of $|\delta \nu_0|$ and $|f_i(0)|$ for $L_F <  \xi_{T_c}$ in Fig.~6 are clearly different from 
those for $x< \xi_{T_c}$ in Fig.~7. In Fig.~6, the Thouless energy $E_\mathrm{Th}$ is larger than the exchange potential $E_\mathrm{ex}$ for
$L_F \ll  \xi_{T_c}$. 
As already mentioned in Sect.~3.1, the effect of the magnetization on the proximity effect is negligible in such a case. 
Therefore all spin-triplet components are absent in the ferromagnet, as shown in Fig.~6, which leads to a minigap structure in LDOS.
On the other hand, in Fig.~7, the choice of $L_F=15 \xi_{T_c}$ means a sufficiently large $E_\mathrm{ex}/E_\mathrm{Th}$ leading to 
the appearance of spin-triplet components near the SF interface. Therefore,  clear minigap structures are not expected 
in experiments by zero-energy peak spectroscopy.
Note that the spatial dependence of the LDOS of an inhomogeneous SF junction for a $weak$-proximity-effect regime and small exchange fields 
has been investigated by Cottet.~\cite{Cottet11}

\section{Conclusions}

To summarize, we have systematically investigated the superconducting proximity effect in SF junctions with  uniform and  nonuniform magnetizations in terms of spin-triplet odd-frequency pairing.
By solving the nonlinear Usadel equation fully numerically, we have calculated the LDOS in a ferromagnet and found following the remarkable results.

(1) In contrast to the case of the uniform magnetization,~\cite{Yokoyama2005,Yokoyama2006} the LDOS in SF junctions with the nonuniform magnetization has a ZEP in a wide range of parameters, indicating {\it the robust presence of the ZEP induced by spin-triplet odd-frequency pairs}.

(2) The ZEP height is damped very slowly with increasing $L_F$ owing to the development of long-range spin-triplet pairing.
This behavior is in marked contrast to the uniform magnetization case in which the zero-energy LDOS shows exponentially damped oscillation as a function of $L_F$.~\cite{Vasenko2011,Buzdin00,Vasenko,kontos2,Boden}

(3) The dip position of $|\delta \nu_0|$ corresponds to the crossover point between singlet and triplet or even- and odd-frequency pairings.
This means that  ZEP spectroscopy can give us clear information on the symmetry of Cooper pairs.

The above remarkable results clearly indicate that the experimental observation of the ZEP for SF junctions 
with a nonuniform magnetization provides the evidence of the existence of the novel spin-triplet odd-frequency pairing. 

In this paper, we have discussed the proximity effect, assuming
a spin-singlet $s$-wave superconductor as a bulk state of S.
An extension to unconventional superconductors is possible
on the basis of more general boundary conditions
~\cite{Proximityd,Proximityd1}
taking the Andreev bound state (ABS)~\cite{ABS1,ABS2} into account.
There have been many studies in various systems 
that show that ABS supports the generation of odd-frequency pairing.~\cite{odd,odd2,odd3,odd4,odd5,odd6,odd7}
The proximity effect in spin-triplet $p$-wave superconductors
is interesting
~\cite{Proximityp,Proximityp1,Proximityp3,Brydon,Brydon2,Higashitani1} since
the odd-frequency pairing induced from
bulk superconductors without exchange energy becomes prominent.~\cite{odd}
In addition, we have particularly focused on the
LDOS. 
It is interesting to discuss the
anomalous Meissner effect~\cite{Proximityp2,Yokoyama2011,Higashitani2}
and surface impedance~\cite{Asano2011} due to the
proximity effect caused by odd-frequency pairing.

\begin{acknowledgments}

%\acknowledgment

We  would like to thank T. Akazaki, A. Cottet, N. Birge, M. Blamire, S. Jiang, S. Kashiwaya, A. S. Vasenko, and T. Yokoyama for useful discussions and comments.
One of us (S. K.) would like to thank the Theory Group of Institut Laue-Langevin for their hospitality during the course of this work.
This work was  supported by the Topological Quantum Phenomena (Nos. 22103002 and 22103005) KAKENHI on Innovative Areas, a Grant-in-Aid for Scientific Research (No. 22710096) from MEXT of Japan, and the JSPS Institutional Program for Young Researcher Overseas Visits.

\end{acknowledgments}

\end{document}